\documentclass{article}
\usepackage{graphicx} 
\usepackage[colorlinks=true, allcolors=blue]{hyperref}
\usepackage[square, sort,comma,numbers]{natbib} 
\usepackage[letterpaper,margin=0.75in]{geometry}
\usepackage{amsmath}

\title{Limitations of detecting structural changes and time-reversal symmetry breaking in scanning tunneling microscopy experiments}
\author{Christopher Candelora and Ilija Zeljkovic}

\begin{document}

\maketitle

The family of kagome superconductors $A$V$_3$Sb$_5$ ($A$=K, Cs, Rb) is an exciting playground for investigating various density waves, unusual superconductivity and a surprising time-reversal symmetry breaking despite the absence of spin magnetism \cite{Wilson2024AV3Sb5Superconductors,Yin2022TopologicalSuperconductors}. The origin of time-reversal symmetry breaking has been of particular interest, and conflicting results have been intensely debated. Scanning tunneling microscopy (STM) provided crucial evidence by the observation that apparent chirality associated with the 2 $\times$ 2 charge density wave (CDW) may be modified by the direction of magnetic field, a phenomenon observed in select sample areas in some experiments \cite{Jiang2021chiral_hasan,Shumiya2021}, but not others \cite{Li2022,Li2022HHWen}. Related to this, Xing \textit{et al.} \cite{xing} investigated the effects of magnetic and electric fields on the 2 $\times$ 2 CDW state and the lattice structure of kagome superconductor RbV$_3$Sb$_5$. They report a $\sim$ 1\% change in the in-plane lattice constants, concomitant with the CDW intensities modification, controlled by the field direction. This was interpreted as a rare case of piezomagnetism. Here we show how the apparent magnetic field induced lattice and CDW intensity change is a consequence of two independent experimental artifacts: reconfiguration of atoms at the STM tip apex that alter the amplitudes of CDW modulations, and piezo creep, hysteresis, and thermal drift that artificially distort STM topographs. We find no evidence supporting that magnetic field leads to intrinsic changes in the sample, which challenges reported piezomagnetism.\\ 

\begin{figure}
    \includegraphics[width = \textwidth]{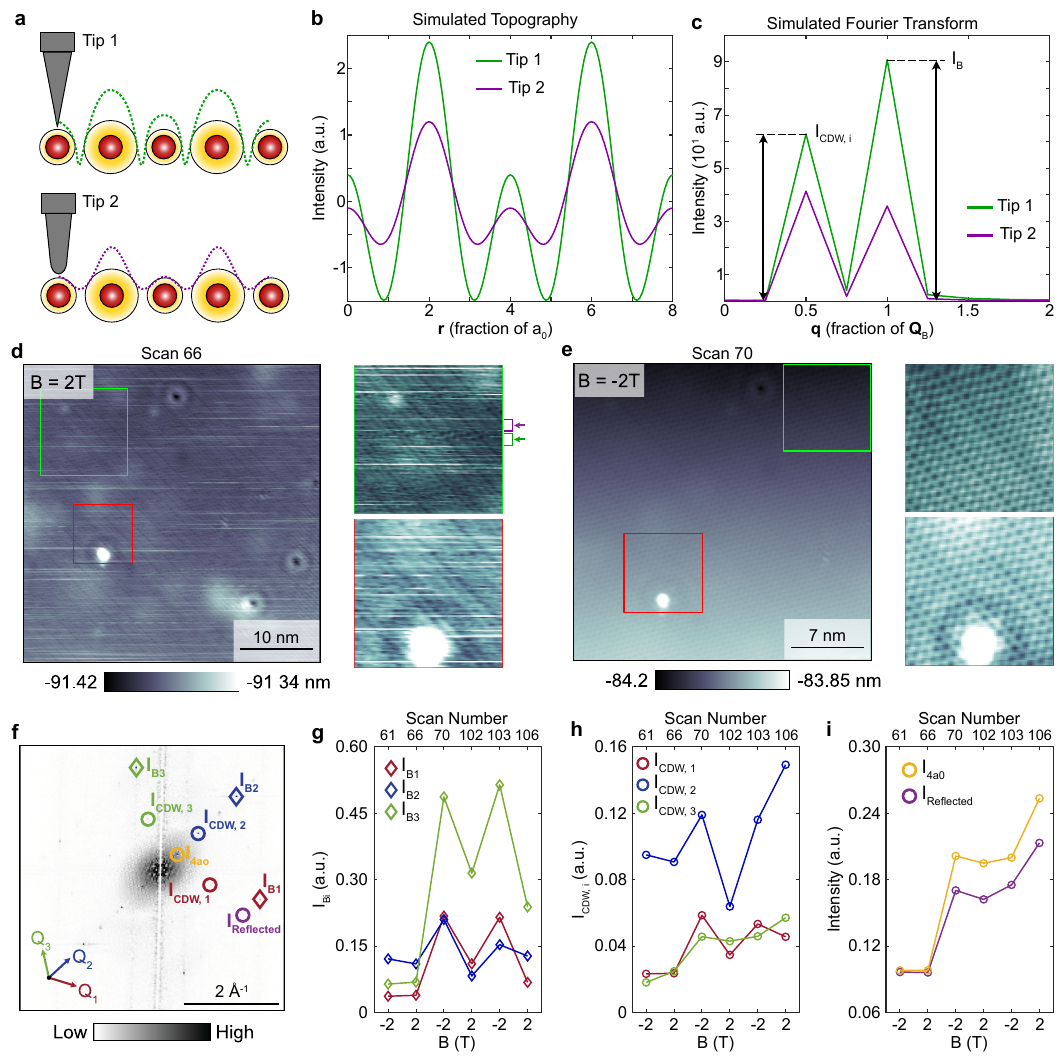}
    \caption{\textbf{Effect of tip changes on the apaprent CDW intensities}. \textbf{a}, A schematic showing the difference in scan motion between an atomically sharp tip (tip 1) and a tip that changed to be slightly more rounded (tip 2) over a 1D lattice with a 2a$_0$ CDW. \textbf{b}, A simulated topographic linecut for the two different tips and \textbf{c}, their corresponding Fourier transform peaks with the CDW intensity (I$_{CDW, i}$) and Bragg intensity (I$_B$) defined. \textbf{d-e}, Raw data of two select STM topographs, the Fourier transform of which were used in Figure 4 of Xing, \textit{et al.} \cite{xing}. Insets to the right show various background subtracted regions (green) and a region common to both topographs (red). The green inset in \textbf{d} shows tip changes that modify the apparent sharpness of the lattice, switching between a duller tip (denoted with purple arrow, similar to tip 2 in \textbf{a}) and a sharper tip (denoted with green arrow, similar to tip 1). The green inset of \textbf{e} shows piezo creep, an artificial distortion of the lattice due to inadequate piezo relaxation. \textbf{f}, Representative Fourier Transform of a topography, with the atomic Bragg peaks denoted by diamonds and CDW peaks denoted by circles. \textbf{g}, Intensities of the atomic Bragg peaks as a function of magnetic field for all 3 lattice directions. \textbf{h}, Intensities of the 2a$_0$ CDW peaks as a function of magnetic field for all 3 lattice directions. \textbf{i}, Intensity of the 4a$_0$ CDW peak and the intensity of a reflected 4a$_0$ CDW peak as a function of magnetic field.}
    \label{fig:1}
\end{figure}

We first discuss the magnetic field experiment, and the apparent change in the CDW intensities in the only dataset presented (Figure 4 in Ref.~\cite{xing}). Visual inspection of raw data (Fig.~\ref{fig:1}d-e, Extended Data Fig.~\ref{ED:1}) already shows obvious differences in the data quality due to the reconfiguration of atoms at the STM tip apex (tip “changes”). These lead to a change in the apparent sharpness and anisotropy of real-space modulations (i.e. atomic corrugations or the CDW signal) detected by the tip (Fig.~\ref{fig:1}a-c). This is also reflected in the change in the intensities of the associated Fourier transform (FT) peaks along different directions. Figure \ref{fig:1}g-i plots the intensities of relevant FT peaks. First, it can be seen that the intensities of \textit{all} FT peaks change in different data sets (Fig.~\ref{fig:1}g-i), not just the 2$a_0$ CDW peaks \textbf{I}$_{CDW,i}$ ($i$=1, 2 or 3). Importantly, atomic Bragg peak intensities change by as much as five-fold (Fig.~\ref{fig:1}g); these should be nearly identical in different topographs without tip changes. This points to a series of tip changes, which are ignored in the work. Second, there are no trends when examining the evolution of individual FT peaks. For instance, \textbf{I}$_{CDW,3}$ can become dimmer (from \#70 to \#102) or brighter (\#103 to \#106) under an identical field change (Fig.~\ref{fig:1}h). The authors only focus on relative intensities, but this does not eliminate tip change effects that are typically anisotropic (i.e.~they change one real-space direction more than another). Therefore, the apparent CDW intensity manipulation is not related to magnetic field change, but instead can be explained by a series of indiscriminate tip changes.\\

\begin{figure}
    \includegraphics[width = \textwidth]{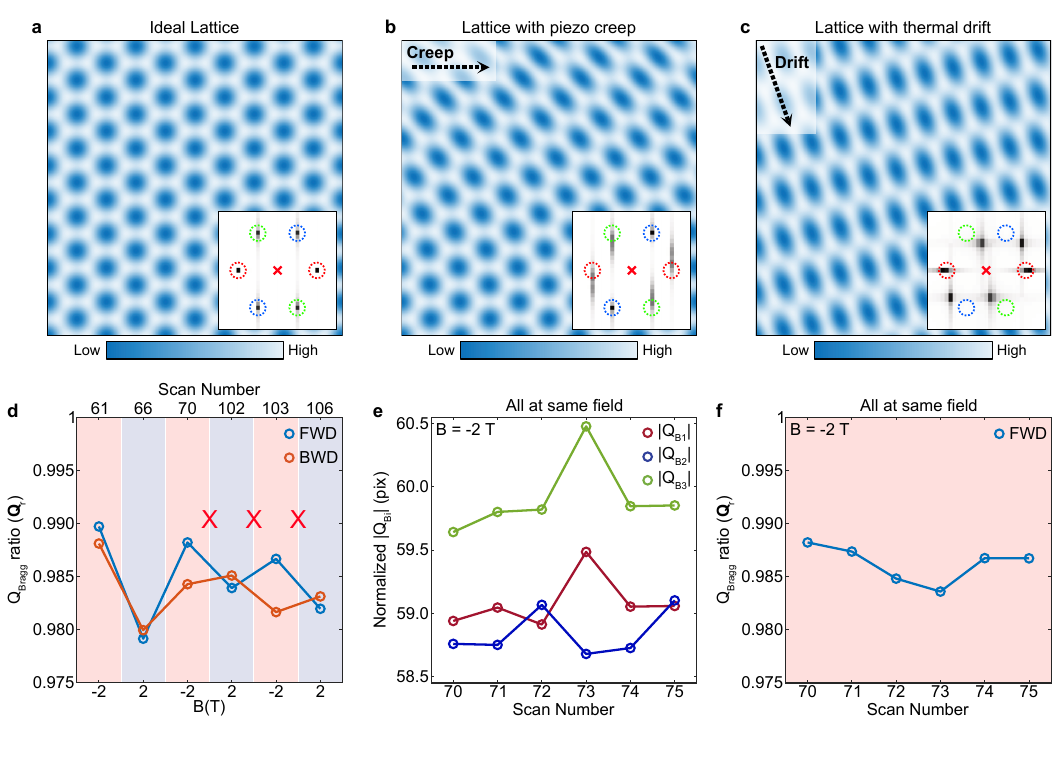}
    \caption{\textbf{Effects of piezo creep and thermal drift on measuring the lattice constants}. \textbf{a-c}, Simulated Sb-terminated surfaces and their corresponding Fourier transforms (inset) for: \textbf{(a)} an ideal lattice without experimental artifacts, \textbf{(b)} with exaggerated piezo creep, and \textbf{(c)} with thermal drift. Note that both artifacts affect the intensity and position of the peaks, with the thermal drift simulated in \textbf{c} shifting the peaks in a similar manner to what was reported by Xing, \textit{et al.} \cite{xing} (Both peaks move down and to the right, putting green now closer to the center and blue farther). \textbf{d}, $Q_{Bragg}$ ratio defined as $|\textbf{Q}_{B1}|$/$|\textbf{Q}_{B3}|$ as a function of magnetic field. The data from the forward scans (blue circles) are shown in Figure 4h of Xing, \textit{et. al.} \cite{xing}. The data from the backwards scans (orange circles) are now presented for comparison. Red X's indicate an inconsistency between the forwards and backwards scans. \textbf{e}, Normalized atomic Bragg vector lengths for the three lattice directions for several consecutive scans, determined via a center-of-mass fitting, all taken at B$_z$ = -2 T. \textbf{f}, $Q_{Bragg}$ ratio for the scans also presented in \textbf{(e)}.}
    \label{fig:2}
\end{figure}

\begin{figure}
    \renewcommand{\thefigure}{1}  
    \renewcommand{\figurename}{Extended Data Figure}  
    \begin{center}
        \includegraphics[width = \textwidth]{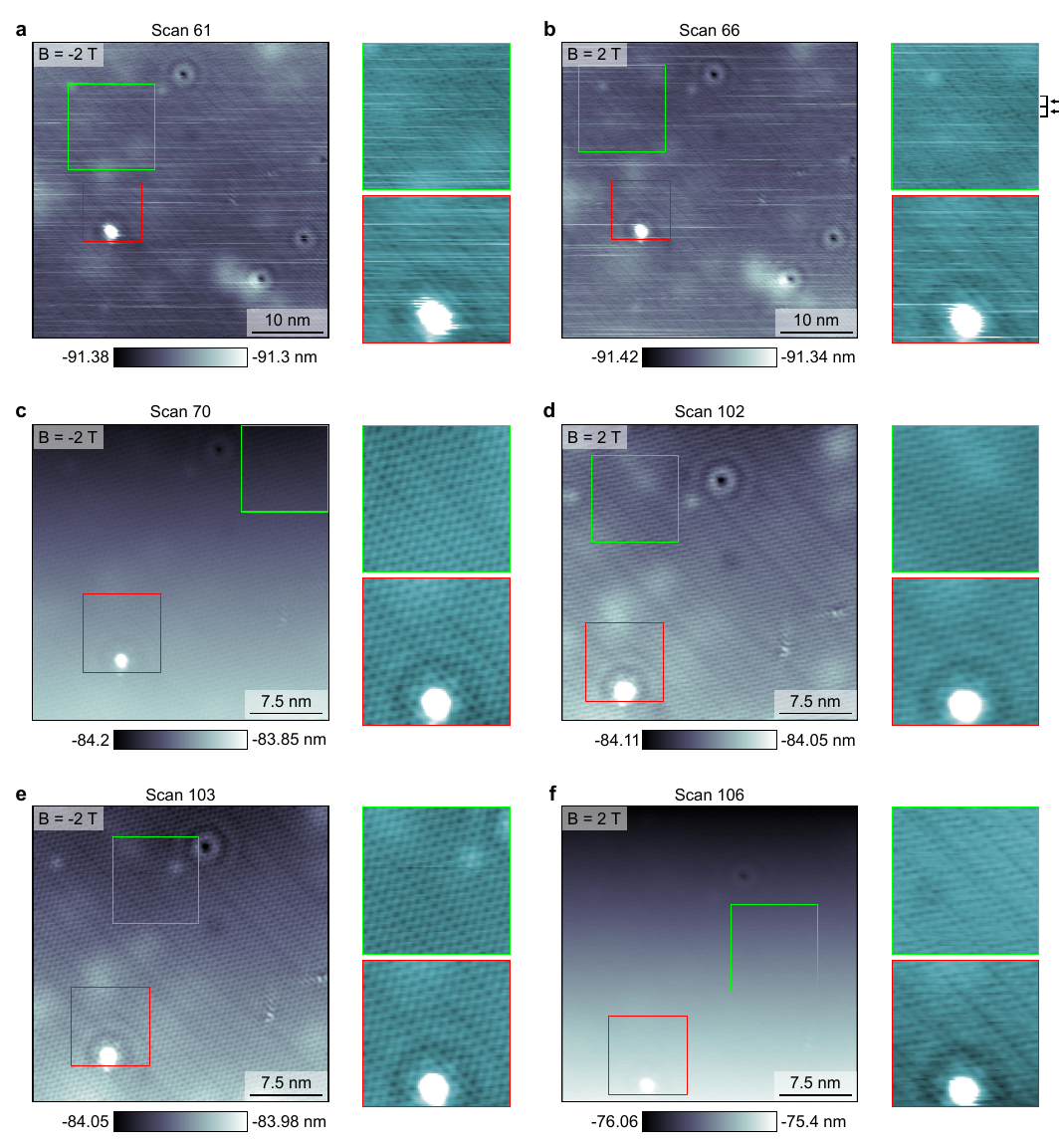}
    \end{center}
    \caption{\textbf{Raw STM topographs}. \textbf{a-f}, Raw STM topographs used for Figure 4 in Xing \textit{et. al.} \cite{xing}. Insets to the right show various background subtracted regions (green) and a region common to all topographs (red). Insets highlight the striking differences in the quality of data at different magnetic fields. The following is an explanation as to why certain areas were circled in green: \textbf{(a)} Shows the instability of the tip; in particular, these horizontal lines generate ''noise" that interferes with detecting the lattice and the CDW signal in the Fourier transform. \textbf{(b)} Area similar to (a), however here one could see how the apparent shape and intensity of the lattice starkly changes as denoted by the arrows and brackets. \textbf{(c)} A rare topograph with a clear lattice and the CDW signal, without striped tip instabilities (albeit with substantial piezo drift as seen from the overall slope in the main topograph in (c), and piezo creep artificially stretching the lattice seen at the top of the green inset due to insufficiently relaxed piezos).}
    \label{ED:1}
\end{figure}

Related to the measurement of the lattice constant change, we note that STM topographs are susceptible to artifacts that distort images: thermal drift, and piezoelectric creep and hysteresis (Fig.~\ref{fig:2}a-c), which were again not considered in Ref.~\cite{xing}. These are generally non-uniform and can evolve as a function of time. Such artifacts can alter the STM-measured atomic lattice constant by a few percent (recent pedagogical example in Ref.~\cite{candelora}). We examine atomic Bragg peak lengths \textbf{Q}$_{B,i}$ ($i$=1, 2 or 3) as a function of magnetic field in detail (Methods), and list major concerns below (additional problems with scan speeds used and reported magnetic fields are in Supplementary Note 1).\\

First, we note that there is no systematic behavior in how individual Bragg peak lengths change. The lack of systematics is summarized by the arrows in Extended Data Fig.~\ref{fig:2}), which show if the apparent \textbf{Q}$_{B,i}$ increased ($\uparrow$) or decreased ($\downarrow$) upon the reversal of field. One would expect a clear $\uparrow$-$\downarrow$-$\uparrow$-etc pattern, but this is not the case. The authors choose to again present the ratio \textbf{Q}$_{B,1}$/\textbf{Q}$_{B,3}$ instead of individual Bragg peak lengths, which ignores these inconsistencies, but as we show next, this approach is still not self-consistent. Second, "control group" data acquired at the same time but not presented in the original paper contradicts the conclusion presented. As the tip rasters across the surface, it first moves from left to right (FWD), then right to left (BWD) before moving to the next line. These yield two nominally equivalent topographs, FWD and BWD, for each scan. Any physical conclusion drawn must be consistent with both. Although \textbf{Q}$_{B,1}$/\textbf{Q}$_{B,3}$ for FWD scans appears to show a systematic 'zig-zag' trend, the BWD scans entirely contradicts the trend for the majority of data points (Fig.~\ref{fig:2}d). This demonstrates that the measurements are not self-consistent, and that measured Bragg peak length changes are influenced by experimental artifacts, not by an intrinsic sample change. The discrepancy cannot be attributed to the size of the field-of-view that is slightly larger for the first two topographs (Supplementary Figure 3). Third, we evaluate consecutive scans under the same magnetic field as another check. It can be seen that individual Bragg peak lengths can vary by up to 2\% in consecutive topographs at same field (Fig.~\ref{fig:2}e,f), highlighting the enormous error of the experimental process. Lastly, we analyze the region where no magnetic field effect is claimed (Extended Data Figure 9 in Ref.\ \cite{xing}). \textbf{Q}$_{B,1}$/\textbf{Q}$_{B,3}$ change for FWD scans is similarly large as that in the area where piezomagnetism is reported and also very different between FWD and BWD scans (Supplementary Information Fig.~2g). This again illustrates the large measurement error that is at least as large as any effect claimed, and further invalidates the authors' argument that always ties the CDW intensity change and the Bragg peak length change.  \\

We conclude that there is no evidence to suggest magnetic field manipulation of the lattice constants. The experiment is not suitably controlled and lacks self-consistency: there is a varying degree of drift (in part seen as the background “slope” in many raw topographs shown in Extended Data Fig.~\ref{ED:1}), which leads to the spurious shift of the atomic Bragg peaks. Such $\sim$1\% shifts can be generated if only single-scans are compared at each field \cite{candelora}, which is the process employed by Ref.~\cite{xing}. There are also concerns related to the reported magnetic field values (Supplementary Note 1). \\

Light manipulation data in Figure 3 of Ref.\ \cite{xing} suffers from much of the same artifacts above (Extended Data Fig.~\ref{ED:3}a-f). There are apparent tip changes from visual examination of data (defects look differently, and there is a varying degree of drift seen from different "slopes" in raw images). Atomic Bragg peak intensities and the \textbf{I}$_{CDW,2}$ again vary by as much as five-fold in an indiscriminate manner between sequences (Extended Data Fig.~\ref{ED:3}). Given that the magnitude of the reported Bragg peak shifts and the CDW intensity changes can be easily generated by tip changes and piezo/thermal drift as we show above, it is unclear if and to what extent there is any electric field induced effect.\\

In summary, the authors ignored tip changes and the effects of piezoelectric and thermal drift when acquiring and analyzing data. We demonstrated an enormous error of their magnetic field measurements, the lack of internal self-consistency or systematic behavior, and entirely opposite conclusions that can be drawn based on the datasets and analysis selected to be shown. As such, we believe that conclusions from the magnetic field experiment involving lattice and CDW intensity change, a crucial piece of piezomagnetism reported, are not supported by data. \\

\noindent{\bf Methods}\\
We extract Fourier transform peak intensities from raw data, using the same single pixel method employed in Ref.~\cite{xing}. To calculate the atomic Bragg peak lengths \textbf{Q}$_{B,i}$ ($i$=1, 2 or 3), we use the center-of-mass method also from Ref.\ \cite{xing}, with a 5 x 5 pixel window centered at the brightest pixel. For magnetic field data analysis, in addition to the analysis of SXM files for scans 61, 66, 70, 102, 103 and 106, we analyzed Excel files provided by the authors between these scans. These however only contained one scan direction and were provided without header information so we could not analyze additional parameters, such as scan speeds and differences between FWD and BWD scans.\\


\noindent{\bf Competing financial interests}\\
The authors declare no competing financial interests.\\

\noindent{\bf Author contributions}\\
C.C. performed the analysis under the supervision of I.Z. C.C. and I.Z. co-wrote the comment.\\

\bibliographystyle{unsrt}
\bibliography{biblio}

\begin{thebibliography}{1}

\bibitem{Wilson2024AV3Sb5Superconductors}
Stephen~D. Wilson and Brenden~R. Ortiz.
\newblock {A}{V}$_{3}${Sb}$_{5}$ kagome superconductors.
\newblock {\em Nature Reviews Materials}, 9(6):420--432, 5 2024.

\bibitem{Yin2022TopologicalSuperconductors}
Jia-Xin Yin, Biao Lian, and M.~Zahid Hasan.
\newblock {Topological kagome magnets and superconductors}.
\newblock {\em Nature}, 612(7941):647--657, 12 2022.

\bibitem{Jiang2021chiral_hasan}
Yu-Xiao Jiang, Jia-Xin Yin, M.~Michael Denner, Nana Shumiya, Brenden~R. Ortiz, Gang Xu, Zurab Guguchia, Junyi He, Md~Shafayat Hossain, Xiaoxiong Liu, Jacob Ruff, Linus Kautzsch, Songtian~S. Zhang, Guoqing Chang, Ilya Belopolski, Qi~Zhang, Tyler~A. Cochran, Daniel Multer, Maksim Litskevich, Zi-Jia Cheng, Xian~P. Yang, Ziqiang Wang, Ronny Thomale, Titus Neupert, Stephen~D. Wilson, and M.~Zahid Hasan.
\newblock Unconventional chiral charge order in kagome superconductor {KV}$_3${S}b$_5$.
\newblock {\em Nature Materials}, 20:1353--1357, 10 2021.

\bibitem{Shumiya2021}
Nana Shumiya, Md~Shafayat Hossain, Jia-Xin Yin, Yu-Xiao Jiang, Brenden~R. Ortiz, Hongxiong Liu, Youguo Shi, Qiangwei Yin, Hechang Lei, Songtian~S. Zhang, Guoqing Chang, Qi~Zhang, Tyler~A. Cochran, Daniel Multer, Maksim Litskevich, Zi-Jia Cheng, Xian~P. Yang, Zurab Guguchia, Stephen~D. Wilson, and M.~Zahid Hasan.
\newblock Intrinsic nature of chiral charge order in the kagome superconductor {RbV}$_3${S}b$_5$.
\newblock {\em Physical Review B}, page 035131, 7.

\bibitem{Li2022}
Hong Li, He~Zhao, Brenden~R Ortiz, Takamori Park, Mengxing Ye, Leon Balents, Ziqiang Wang, Stephen~D Wilson, and Ilija Zeljkovic.
\newblock Rotation symmetry breaking in the normal state of a kagome superconductor {KV}$_3${S}b$_5$.
\newblock {\em Nature Physics}, pages 265--270, 3.

\bibitem{Li2022HHWen}
Huazhou Li, Siyuan Wan, Han Li, Qing Li, Qiangqiang Gu, Huan Yang, Yongkai Li, Zhiwei Wang, Yugui Yao, and Hai-Hu Wen.
\newblock No observation of chiral flux current in the topological kagome metal {CsV}$_3${S}b$_5$.
\newblock {\em Physical Review B}, page 045102, 1.

\bibitem{xing}
Y.~Xing, S.~Bae, E.~Ritz, F.~Yang, T.~Birol, A.N.C. Salinas, B.R. Ortiz, S.D. Wilson, Z.~Wang, R.M. Fernandes, and V.~Madhavan.
\newblock Optical manipulation of the charge-density-wave state in {R}b{V}$_3${S}b$_5$.
\newblock {\em Nature}, \textbf{631}, 60-66, (2024).

\bibitem{candelora}
C.~Candelora, H.~Li, , M.~Xu, B.R. Ortiz, A.N.C. Salinas, S.~Cheng, A.~LaFleur, Z.~Wang, S.D. Wilson, and I.~Zeljkovic.
\newblock Quantifying magnetic field driven lattice distortions in kagome metals at the femtometer scale using scanning tunneling microscopy.
\newblock {\em Phys. Rev. B}, \textbf{109}, 155121, (2024).

\end{thebibliography}

\newpage

\clearpage

\clearpage

\clearpage

\begin{figure}
    \renewcommand{\thefigure}{2}  
    \renewcommand{\figurename}{Extended Data Figure} 
    \centering
    \includegraphics[width = \textwidth]{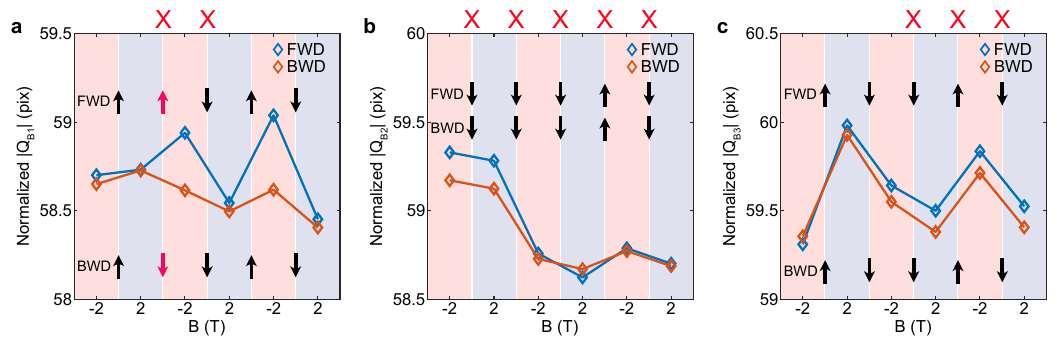}
    \caption{\textbf{Analysis of Bragg lengths as a function of magnetic field}. \textbf{a-c}, Normalized Bragg lengths as a function of magnetic field from the topographs used in Figure 4 of Xing \textit{et al.} \cite{xing} along $Q_{B1}$, $Q_{B2}$, and $Q_{B3}$, respectively. Arrows indicate the direction of change in intensity for the forward scans (top row of arrows) and backward scans (bottom row of arrows), with red arrows indicating an inconsistency between forwards and backwards scans. Red X's above indicate inconsistencies between the proposed switching (that is, $|Q_{B1}|$ decreasing and $|Q_{B3}|$ increasing while $|Q_{B2}|$ remains constant for $B_z$ $>$ 0, and the opposite for $B_z$ $<$ 0) and the data.}
    \label{ED:2}
\end{figure}

\clearpage

\begin{figure}
    \renewcommand{\thefigure}{3}  
    \renewcommand{\figurename}{Extended Data Figure} 
    \centering
    \includegraphics[width = \textwidth]{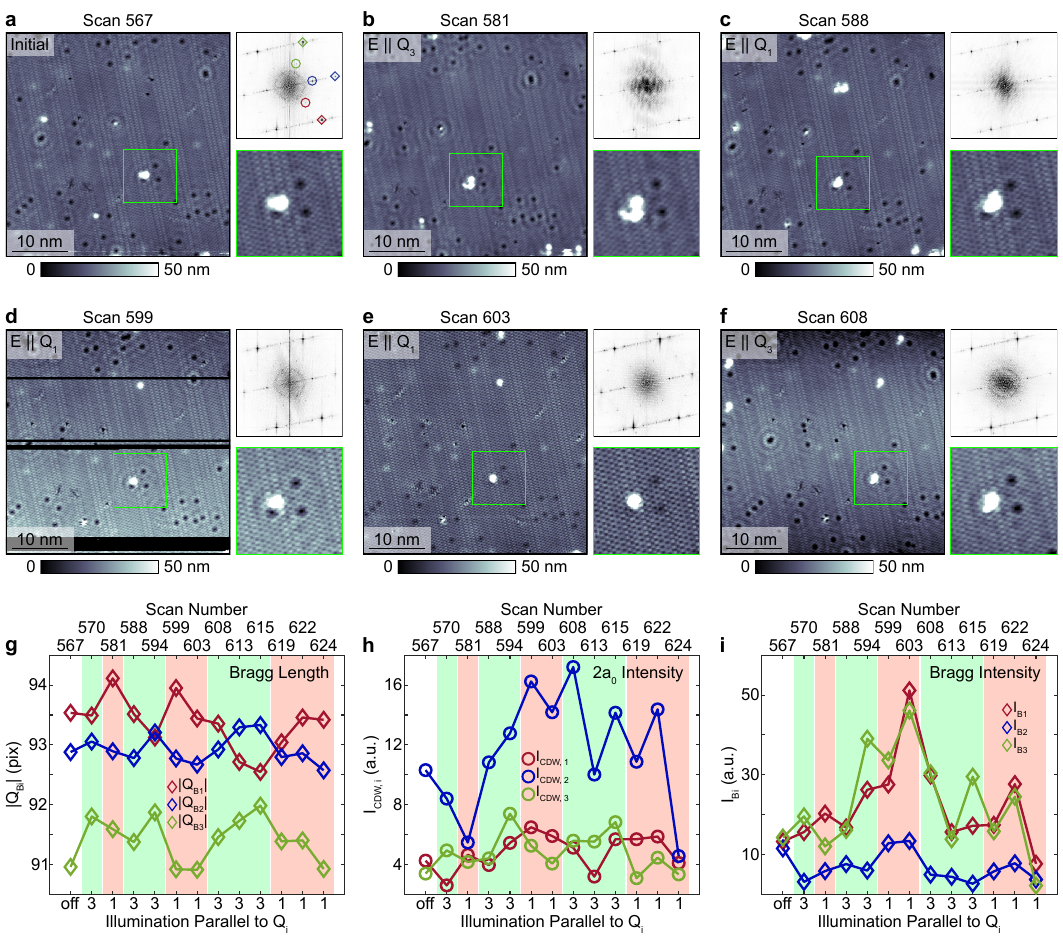}
    \caption{\textbf{Analysis of laser data}. \textbf{a-f}, A sample of six plane-subtracted STM topographs used for Figure 3 in Xing \textit{et al.} \cite{xing}. Insets to the right show their Fourier transform (top inset) as well as a region common to all topographs (bottom inset). Fourier transform in \textbf{(a)} indicates CDW peaks by circles and Bragg peaks by diamonds, with $Q_1$, $Q_2$, and $Q_3$ directions represented by red, blue, and green, respectively. \textbf{(a)} Shows the first scan used as the "inital condition." \textbf{(b-c)} show the first transition going from illumination along $Q_3$ to along $Q_1$. \textbf{(d-e)} two topographs taken successively along $Q_1$, and \textbf{(f)} switching back to $Q_3$ illumination. Notice how going from \textbf{(a)} to \textbf{(b)} the tip becomes double, as apparent from the QPI ring disappearing in the Fourier transform (top insets) and the number of impurities doubling (bottom insets, though present throughout topograph \textbf{(b)}). The tip is changed again going from \textbf{(b)} to \textbf{(c)} as shown by the change in shape of the impurities (bottom inset). Notice also the qualitative differences of \textbf{(d)} and \textbf{(e)}, which were taken with the same illumination direction $Q_1$. The QPI ring again disappears between the two Fourier transforms, and the vacancies appear glaringly different (bottom insets). \textbf{g}, Bragg lengths along $Q_1$ (red), $Q_2$ (blue), and $Q_3$ (green) as a function of laser illumination direction from topographs used in Figure 3 of Xing \textit{et al.} \cite{xing}. Lengths were determined by 5 x 5 COM fitting of the Bragg peaks. \textbf{h}, CDW intensity along along $Q_1$ (red), $Q_2$ (blue), and $Q_3$ (green) as a function of laser illumination direction acquired by using the Fourier transform of the raw data. \textbf{i}, Bragg intensity along $Q_1$ (red), $Q_2$ (blue), and $Q_3$ (green) as a function of laser illumination direction acquired by analyzing the Fourier transform of the raw data. Note the large variations in the I$_{CDW,2}$ in (h) and atomic Bragg peak intensities in (i). }
    \label{ED:3}
\end{figure}

\end{document}